\documentclass[11pt,letterpaper]{article}

\usepackage[T1]{fontenc}
\usepackage{newtxtext}
\usepackage[margin=1in,headheight=22pt]{geometry}
\usepackage{microtype}
\usepackage{xcolor}
\definecolor{accent}{HTML}{496778}
\definecolor{ink}{HTML}{292824}
\definecolor{muted}{HTML}{77746E}
\definecolor{rule}{HTML}{D9D5CE}
\definecolor{abstractbg}{HTML}{F2F5F6}
\usepackage{amsmath,amssymb,amsthm}
\usepackage{stmaryrd}
\usepackage{graphicx}
\usepackage{tikz}
\usepackage{wrapfig}
\usepackage{algorithm}
\usepackage[noEnd=true,indLines=true,commentColor=black]{sty/algpseudocodex}
\usepackage{booktabs}
\usepackage{subcaption}
\usepackage{siunitx}
\usepackage{makecell}
\usepackage{multirow}
\usepackage{newfloat}
\usepackage{listings}
\usepackage[numbers,sort&compress]{natbib}
\usepackage{titlesec}
\usepackage{fancyhdr}
\usepackage[most]{tcolorbox}
\usepackage[hidelinks]{hyperref}
\usepackage{cleveref}
\usepackage{sty/custom}
\hypersetup{colorlinks=true,linkcolor=accent,citecolor=accent,urlcolor=accent}

\titleformat{\section}
  {\Large\bfseries\color{ink}}{\color{ink}\thesection}{0.65em}{}
\titleformat{\subsection}
  {\large\bfseries\color{ink}}{\color{ink}\thesubsection}{0.65em}{}
\titleformat{\subsubsection}
  {\normalsize\bfseries}{\thesubsubsection}{0.65em}{}
\titlespacing*{\section}{0pt}{2.2ex plus 0.5ex minus 0.2ex}{0.8ex}
\titlespacing*{\subsection}{0pt}{1.8ex plus 0.4ex minus 0.2ex}{0.5ex}

\renewcommand{\headrulewidth}{0.35pt}
\renewcommand{\headrule}{\hbox to\headwidth{\color{rule}\leaders\hrule height \headrulewidth\hfill}}

\newtcolorbox{abstractbox}{
  enhanced,
  breakable,
  colback=abstractbg,
  colframe=abstractbg,
  boxrule=0pt,
  borderline west={2pt}{0pt}{accent},
  sharp corners,
  left=11pt,
  right=11pt,
  top=9pt,
  bottom=9pt,
  before skip=0.3em,
  after skip=0.6em
}
\renewenvironment{abstract}{%
  \begin{abstractbox}
  \setlength{\parindent}{0pt}%
  {\small\bfseries\color{accent}\MakeUppercase{Abstract}}\par\smallskip
  \small
}{%
  \end{abstractbox}
}

\DeclareCaptionStyle{ruled}{labelfont=normalfont,labelsep=colon,strut=off}
\floatstyle{ruled}
\newfloat{listing}{tb}{lst}{}
\floatname{listing}{Listing}

\newlength{\preprintlabelwidth}
\title{%
  \begin{minipage}{\textwidth}\raggedright
  \setlength{\parskip}{0pt}%
  \settowidth{\preprintlabelwidth}{\scriptsize PREPRINT}%
  \noindent{\color{accent}\vbox{%
    \hbox{\scriptsize PREPRINT}%
    \kern 2pt%
    \hrule width \preprintlabelwidth height 0.4pt%
  }}\par\vspace{1em}%
  {\LARGE\bfseries\color{ink} Solving Multi-Agent Sokoban via LaCAM}%
  \end{minipage}%
}
\author{%
  \begin{minipage}{\textwidth}\raggedright
  {\normalsize\bfseries\color{ink} Keisuke Okumura}\\[0.15em]
  \small\textcolor{muted}{National Institute of Advanced Industrial Science and Technology (AIST), Japan}\\[-0.2em]
  \small\href{mailto:okumura.k@aist.go.jp}{okumura.k@aist.go.jp}
  \end{minipage}%
}
\date{}

\makeatletter
\renewcommand{\maketitle}{%
  \thispagestyle{plain}%
  \vspace*{0.4em}%
  \noindent\@title\par
  \vspace{0.3em}%
  \noindent\@author\par
  \vspace{0.2em}%
}
\makeatother

\begin{document}
\maketitle
\begin{abstract}
Sokoban, a puzzle game in which an agent pushes boxes onto unlabelled target locations in a grid world, is a long-standing benchmark planning problem.
While it is easy to see the connection to practical applications such as warehouse logistics with autonomous forklifts, its multi-agent counterpart has remained underdeveloped.
This is because Multi-Agent Sokoban is substantially more difficult due to factors specific to multi-agent planning, such as the rapidly growing branching factor as the number of agents grows and the need to handle integrated task assignment and collision-free pathfinding.
In this paper, we show that a scalable planner for Multi-Agent Sokoban can be designed by leveraging recent advances in multi-agent pathfinding (MAPF).
Specifically, our Sokoban-LaCAM efficiently solves instances involving tens of agents and boxes while preserving both completeness and eventual optimality guarantees.
This provides evidence that MAPF can serve as a powerful primitive for solving broader collective automation problems.
\end{abstract}

\noindent{\small\textcolor{muted}{Project page: }\url{https://kei18.github.io/multi-agent-sokoban/}}
\par
\section{Introduction}
Sokoban is a popular puzzle game in which an agent pushes boxes onto target locations in a grid world, but also a long-standing benchmark problem for heuristic search due to its computational difficulty~\cite{dor1999sokoban,junghanns2001sokoban,shoham2021solving}.
For example, when a box is adjacent to a wall, the agent cannot push it from that direction, restricting the set of reachable box configurations.
This asymmetry gives rise to dead-end situations, leading search algorithms to spend substantial effort exploring fruitless branches in the absence of effective heuristics, which are notoriously difficult to design.
Not surprisingly, Sokoban is PSPACE-complete from the perspective of computational complexity~\cite{culberson1997sokoban}.

By contrast, its multi-agent counterpart, which we refer to as \emph{Multi-Agent Sokoban}, has received little attention.
Only a few papers have explicitly considered this problem~\cite{lawton2008multi,zhang2009multi}, and no dedicated solver is publicly available to the best of our knowledge.
Nevertheless, Multi-Agent Sokoban poses interesting challenges in multi-agent planning while maintaining strong connections to real-world problems.
For example, a planner must determine both which agent should transport each box and which target location each box should be assigned to.
It must then compute collision-free paths while avoiding agent--agent, agent--box, and box--box collisions, as well as dead-end situations.
This structure closely resembles warehouse logistics~\cite{wurman2008coordinating}, where multiple autonomous robots (e.g., AGVs and forklifts) transport goods (e.g., parcels and pallets) to designated locations while coordinating their movements to avoid mutual interference and maximise overall throughput.

We therefore regard Multi-Agent Sokoban as a fruitful benchmark for multi-agent planning, where a planner needs to address integrated task assignment and coordinated path planning.
In particular, we ask whether \emph{multi-agent pathfinding (MAPF)} techniques can provide a scalable and real-time planning primitive beyond the canonical MAPF formulation.

MAPF~\cite{stern2019def}, the problem of computing collision-free paths for multiple agents on a graph, has been extensively studied since the early 2010s, driven by the increasing demand for coordinated robots in factory automation.
Among recent advances, \emph{LaCAM$^{(\ast)}$}~\cite{okumura2023lacam,okumura2023improving} demonstrates that reliable sub-second planning for hundreds of agents is possible.
Since LaCAM searches directly over the configuration space, where a configuration represents the states of all agents, its search framework naturally extends to collective automation problems that admit a configuration-space formulation.

To this end, this paper presents \emph{Sokoban-LaCAM} for Multi-Agent Sokoban.
It performs search over the configuration space by generating joint actions via constrained \emph{PIBT}~\cite{okumura2022priority}, where each agent's action is determined based on target--box and agent--box assignments.
To further speed up the solution discovery, we introduce problem-specific heuristics and pruning rules on top of the joint search.
Theoretically, Sokoban-LaCAM is complete and eventually optimal with respect to cumulative transition cost (e.g., plan length; aka. makespan), meaning that it finds an optimal solution in finite time whenever one exists and otherwise correctly reports infeasibility.
Empirically, Sokoban-LaCAM solves instances with tens of agents and boxes within seconds.
This demonstrates substantially better scalability than adapting \ao{IDA}~\cite{korf1985depth}, which has traditionally been used for single-agent Sokoban.

Importantly, the broader goal of this work is not merely to solve Multi-Agent Sokoban, but to demonstrate that search-based MAPF algorithms, at least LaCAM, can serve as general multi-agent combinatorial search frameworks for collective automation problems.
This perspective opens the door to integrating learned heuristics to guide search efficiently~\cite{jain2026graph}, as well as developing effective task-specific configuration generators that exploit the underlying problem structure~\cite{fu2026symbolic,suzuki2026polynomialtime}.
In the remainder of the paper, we first formulate Multi-Agent Sokoban and review related work, with a particular focus on connections to MAPF.
We then present Sokoban-LaCAM and evaluate its performance experimentally.
Throughout the paper, we use $\bot$ as an ``undefined'' sign for convenience.

\section{Preliminaries}
\subsection{Problem Definition}
\label{sec:problem-definition}

Let $G=(V, E)$ be an undirected graph representing a grid environment, where $V$ is the set of traversable cells and $E$ connects orthogonally adjacent cells.
We consider a set of agents $A=\{a_1,\ldots,a_n\}$, a set of boxes $B=\{b_1,\ldots,b_m\}$, and a set of target locations $T\subseteq V$.
We assume that the number of boxes equals the number of targets (i.e., $m=|T|$).

A \emph{configuration} is defined as an injective function $Q: A\cup B\mapsto V$, assigning every \emph{object} (either agent or box) to a distinct vertex.
A \emph{Multi-Agent Sokoban instance} is then defined as
$\langle G, A, B, T, Q\init\rangle$,
where $Q\init$ denotes the initial configuration.
A \emph{solution} is a finite sequence of configurations
$\pi=(Q_0=Q^{\mathrm{init}},Q_1,\ldots,Q\suf{last})$
such that every target vertex is occupied by a box in the final configuration, namely,
\[
\forall g\in T,\ \exists b\in B,\ Q\suf{last}(b)=g.
\]
Furthermore, every (ordered) pair of consecutive configurations must satisfy the transition rules defined below.

A transition from a configuration $Q\from$ to another one $Q\to$ is valid if the following conditions hold:
\emph{(i)}~Every object either stays at its current vertex or moves to an adjacent vertex;
\emph{(ii)}~No swap conflict occurs between any pair of objects, i.e., $\forall i\neq j \in A \cup B, (Q\from(i),Q\from(j)) \neq (Q\to(j),Q\to(i))$;
\emph{(iii)}~For every moved box, there exists exactly one agent occupying the adjacent vertex in the pushing direction that moves into the box's previous location, while the box simultaneously moves one step further in the same direction; and
\emph{(iv)} An agent cannot push two boxes simultaneously.
Note that the vertex conflict constraint is already implied by the definition of a configuration.

Then, the quality of a solution $\pi$ is evaluated by its length, referred to as the \emph{makespan}.
Our objective is to compute a makespan-minimising solution as quickly as possible.

Compared to the standard single-agent formulation, which has been widely studied in the search community~\cite{junghanns2001sokoban,shoham2021solving} and, more recently, in the learning community~\cite{feng2020novel,shoham2021solving}, Multi-Agent Sokoban additionally requires reasoning about inter-agent collisions.
The principal challenge, however, is the dramatically increased branching factor, which grows exponentially with the number of agents $n$ and renders conventional search schemes impractical at scale.
This motivates us to leverage algorithmic insights established in MAPF.

\subsection{MAPF, LaCAM, and PIBT}
\begin{wrapfigure}{R}{0.5\textwidth}
\begin{minipage}{\linewidth}
\newcommand{\parent}{\m{\mathit{parent}}}
\vspace{-3em}
\begin{algorithm}[H]
\small
\caption{LaCAM}
\label{algo:lacam}
\begin{algorithmic}[1]
\Input{MAPF instance $(G, A, Q\init, Q\goal)$}
\Output{solution or \nosolution}
\State initialize \open, \explored
\State $N\init \leftarrow \begin{aligned}[t]\langle
&Q: Q\init,
\tree: \llbracket~[\bot]~\rrbracket,
\parent: \bot
\rangle\end{aligned}$
\label{algo:lacam:init-node}
\State $\open.\push(N\init)$;~~$\explored[Q\init] = N\init$
\While{$\open \neq \emptyset$}
\State $N \leftarrow \open.\funcname{top}()$
\IfSingle{$N.Q = Q\goal$}{\Return $\funcname{backtrack}(N)$}
\IfSingle{$N.\tree = \emptyset$}{$\open.\pop()$;~\Continue}
\smallskip
\Statex {\footnotesize \emph{(lazy constraints addition)}}
\State $C \leftarrow N.\tree.\pop()$
\label{algo:lacam:get-constraint}
\If{$|C| \leq |A|$}
\State $a_i \gets a_{|C|}$;\; $v \gets N.Q[a_i]$
\For{$u \in \neigh(v) \cup \{ v \}$}
\State $N.\tree.\push([C + \langle a_i, u\rangle])$
\EndFor
\EndIf
\label{algo:lacam:constraints-addition:end}
\smallskip
\Statex \quad{\footnotesize \emph{(configuration generation)}}
\State $Q\new \leftarrow \funcname{configuration\_generator}(N.Q, C)$
\label{algo:lacam:config-generator}
\If{$Q\new \neq \bot \land Q\new \not\in \explored$}
\label{algo:lacam:check}
\State $N\new \leftarrow \begin{aligned}[t]\langle&
  Q: Q\new,~
  \tree: \llbracket~[\bot]~\rrbracket,
  \parent: N
  \rangle\end{aligned}$
\State $\open.\push(N\new)$;\;$\explored[Q\new] = N\new$
\EndIf
\EndWhile
\State \Return \nosolution
\label{algo:lacam:unsolvable}
\end{algorithmic}
\end{algorithm}
\end{minipage}
\end{wrapfigure}

The canonical MAPF formulation can be viewed as a special case of Multi-Agent Sokoban by setting $B=\emptyset$ and introducing a goal configuration $Q\goal$, where each agent $a_i$ is required to reach its assigned target $Q\goal(a_i)$.
Motivated by applications in warehouse logistics, numerous solution paradigms have been developed for MAPF, including rule-based methods, dedicated search algorithms, and learning-based approaches~\cite{ma2022graph,alkazzi2024a}, to achieve scalable real-time planning.
One particularly successful paradigm is configuration generator-based search, exemplified by \emph{LaCAM}~\cite{okumura2023lacam}.

\Cref{algo:lacam} presents the pseudocode of LaCAM.
It maintains an \open stack of \emph{search nodes} and an \explored table recording previously visited configurations.
LaCAM expands one search node at a time in a depth-first manner.
As in conventional graph search, each node corresponds to a configuration and stores a pointer to its parent.
In addition, each node maintains a \tree queue of \emph{constraints} used for successor generation (\cref{algo:lacam:init-node}).
Each constraint $C$ specifies which agents take which actions, partially determines the successor configuration (\cref{algo:lacam:get-constraint}--\ref{algo:lacam:constraints-addition:end}).
Then, given the current configuration and a selected constraint, \emph{configuration generator} produces exactly one successor configuration $Q\new$ satisfying $C$ (\cref{algo:lacam:config-generator}).
Rather than explicitly enumerating all feasible successor configurations, whose number grows exponentially with $n$, LaCAM incrementally refines the constraints to lazily explore the successor space.
This postpones the exponential branching inherent in multi-agent planning while preserving the exhaustive search structure, thereby maintaining completeness and enabling highly scalable planning.
Furthermore, a minor modification to \cref{algo:lacam} by introducing search-tree rewiring yields \emph{\ao{LaCAM}}~\cite{okumura2023improving}, an anytime algorithm that eventually converges to an optimal solution.

LaCAM's performance largely depends on the underlying configuration generator, which is typically instantiated by \emph{PIBT}~\cite{okumura2022priority}.
Given, for each agent $a_i$, an ordered list of candidate actions $\phi_i$ called \emph{preference}, PIBT synthesises a feasible successor configuration in time linear in $n$ while guaranteeing a collision-free transition.
This scheme has subsequently been adapted to a range of planning settings beyond the canonical MAPF formulation~\cite{okumura2021timeindependent,yukhnevich2026enhancing,moldagalieva2026db,nagai2026from}, demonstrating the versatility of PIBT as a primitive for scalable multi-agent coordination.

Importantly, neither LaCAM nor PIBT is tied to the canonical MAPF formulation.
Rather, they provide a generic framework for systematic search over configurations, where successor configurations are generated from agent-wise action preferences.
This abstraction has enabled LaCAM to serve as a planning primitive well beyond MAPF, such as aggressive navigation with physical robot fleets~\cite{okumura2026concrete}.
In this work, we take this idea one step further by tackling the substantially more challenging Multi-Agent Sokoban problem, where path planning is tightly coupled with \emph{task planning}, as well as workspace reconfiguration via \emph{object manipulation}, since box movements dynamically alter the traversable workspace.

\subsection{Integrated Task and Path Planning}
In the context of MAPF, task planning is typically formulated as target assignment.
While this problem has long been studied as an independent problem~\cite{gerkey2004formal}, the advent of MAPF has motivated research on integrated multi-agent target assignment and path planning.
One representative formulation is \emph{unlabelled MAPF}, in which interchangeable agents are required to occupy a given set of target locations, while \emph{MAPF-UA} (with unassigned agents)~\cite{felner2026mapfua} considers settings in which a subset of agents may terminate at arbitrary vertices.
A more general formulation is \emph{target assignment and pathfinding (TAPF)}, where an agent--vertex matrix specifies the admissible targets for each agent.
Various algorithms have been developed for unlabelled MAPF~\cite{yu2013multi,okumura2023solving}, MAPF-UA~\cite{fu2026symbolic,gabay2026reachability}, and TAPF~\cite{ma2016optimal,tang2024ita,kumagai2026alternating}, ranging from optimal algorithms to highly scalable suboptimal methods.
Notably, several recent scalable algorithms for these formulations build upon LaCAM, motivating us to extend the framework to Multi-Agent Sokoban.

\subsection{Planning with Object Manipulation}
Sokoban naturally combines path planning with object manipulation, where moving objects continuously modify the traversable workspace for both agents and other objects.
Manipulation planning in which the final obstacle configuration is not part of the planning objective has been studied in both discrete~\cite{demaine2000pushpush,ren2025search} and continuous domains~\cite{wilfong1988motion,stilman2008planning}, as well as in multi-agent settings~\cite{renault2024multi,hu2025conflict}.
In parallel, inspired by warehouse automation, several MAPF variants have also considered transporting shelves or cargo~\cite{vainshtain2021multi,li2023double,bachor2023multi,makino2024marpf,sherma2025agent}, under the assumption that the transporting agent and the transported object are allowed to occupy the same location.
In contrast, Sokoban’s push-only dynamics make dead-ends particularly severe and often irreversible, unlike formulations that permit objects to be carried, co-located with agents, or repositioned more freely.

\section{Sokoban-LaCAM}
\label{sec:sokoban}

This section first presents the core design of Sokoban-LaCAM, followed by the task assignment heuristics that enable it to efficiently solve large-scale instances.
Our objective is not to provide a complete engineering blueprint, but rather to demonstrate that a configuration-based search framework for MAPF can be readily extended to Multi-Agent Sokoban with only modest modifications.
Accordingly, several implementation details are omitted for brevity.

\subsection{Minimal Adaptation}
Our Sokoban-LaCAM directly applies \cref{algo:lacam} to the configuration space of Multi-Agent Sokoban, with a dedicated PIBT-based configuration generator (\cref{algo:lacam:config-generator}) that accounts for box-aware coordination.
Although a configuration is defined by the placements of both agents and boxes, boxes never move independently.
Thus, LaCAM's constraints only need to regulate agent actions.
This results in an exhaustive search regardless of the underlying configuration generator, even one based on random walks, following the analysis in~\cite{okumura2023lacam}.
Consequently, the algorithm is complete; that is, it finds a solution in finite time whenever one exists and otherwise proves that the instance is infeasible.
Moreover, its eventual-optimal variant, \ao{LaCAM}, is readily applicable, yielding an eventually optimal solver for Multi-Agent Sokoban under cumulative cost objectives such as makespan.

The adaptation of PIBT, hereafter referred to as \emph{Sokoban-PIBT}, is likewise straightforward.
Recall that PIBT synthesises a collision-free action for each agent $a_i$ given its preference $\phi_i$, which is an ordered list of applicable actions.
We retain this structure; that is, Sokoban-PIBT only generates agent actions, while the resulting box placements are determined implicitly by these actions.
The only modification is the collision model.
If an agent action pushes an adjacent box, Sokoban-PIBT performs collision checking after simulating the corresponding box movement.
Any action that results in a box--box or box--agent collision is rejected.
Note that this modification does not guarantee that Sokoban-PIBT always produces a feasible configuration under the given constraint.
However, this is not problematic, as such failures are naturally handled by LaCAM, which simply continues searching over alternative constraints.

The remaining challenge is therefore to construct the preferences $\phi_i$ supplied to Sokoban-PIBT, as described next.

\subsection{Preference Construction}
The performance of Sokoban-LaCAM largely depends on the quality of the constructed preferences.
In particular, preference construction must be computationally efficient, as it is invoked at every search iteration, while also providing effective guidance to avoid exploring fruitless branches.
With these considerations in mind, our approach consists of the following two steps for a given configuration $Q$:
\begin{enumerate}
\item \textbf{Target--Box Assignment}:
  We first solve a maximum-cardinality matching problem to determine which box should be assigned to which target, yielding an assignment function $\mtb: T \mapsto B$.
  For each box $b_j$ not already located on its assigned target $t_j \in T$, this assignment immediately determines a \emph{stand} $s_j \in V$, namely the position from which an agent should push $b_j$ towards $t_j$.
  Note that a single box may admit multiple stands because the direction from which it can be pushed towards its target is not necessarily unique.
  Conversely, a box may have no feasible stand because all candidate stands are blocked by other boxes.
\item \textbf{Agent--Stand Assignment}:
  We next solve a matching problem between agents and the identified stands.
  This yields an assignment function $\mas: A \mapsto S \cup \{\bot\}$, where $S$ denotes the set of stands.
  Unlike the target--box assignment, some agents may remain unassigned (mapped to $\bot$), and some stands may not be assigned to any agent (e.g., when $|A| \ll |B|$).
\end{enumerate}
Once these assignments have been made, we derive the preference $\phi_i$ for each agent according to the following rules.

\begin{enumerate}
\item If $\mas(a_i)=s_j$ and $a_i$ is on $s_j$, the box-push action is available.
  Thus, $a_i$ prioritises the action towards $Q(b_j)$;
  the remaining actions are ordered randomly.
\item If $\mas(a_i)=s_j$ but $a_i$ has not yet reached $s_j$, the applicable actions are ordered in ascending order of $\dist_{G \setminus B}(Q(a_i), s_j)$, where $\dist_{G\setminus B}$ denotes the shortest-path distance on the graph induced by removing all box locations, since agents cannot traverse them.
\item When $\mas(a_i) = \bot$, the applicable actions are ordered in ascending order of the resulting value of $\min_{s_j \in S} \dist_{G \setminus B}(Q(a_i), s_j)$.
  This distance is efficiently computed by a multi-source BFS from all stands.
  This encourages idle agents to remain close to potential future tasks.
\end{enumerate}
Together, these rules guide agents towards useful box-pushing actions while naturally distributing work across multiple agents, thereby increasing concurrency and contributing to minimising the plan makespan.
We now describe each assignment procedure in detail.

\subsection{Target--Box Assignment}
Before entering the LaCAM search iterations, Sokoban-LaCAM precomputes a push-aware distance $\dist_p(t, v)$ from each target $t \in T$ to every location $v \in V$.
This distance is computed by a backward BFS rooted at each target while excluding positions from which a box cannot be pushed, such as locations adjacent to walls.
We then seek a target--box assignment \mtb that minimises the maximum push-aware target--box distance,
$\max_{t \in T} \dist_p(t, \mtb(t))$,
as a surrogate objective for reducing the final makespan.
This naturally gives rise to a bottleneck assignment problem~\cite{gross1959bottleneck}.
Instead of solving this problem optimally, however, we adopt a lightweight heuristic, since the final makespan also depends on the subsequent agent--stand assignment and collision-free pathfinding.

Specifically, inspired by a suboptimal unlabelled MAPF algorithm~\cite{okumura2023solving}, we first compute an unweighted bipartite matching between targets and boxes using the Ford--Fulkerson algorithm based on DFS augmenting paths~\cite{ford1956maximal}.
Each augmenting-path search visits candidate boxes in ascending order of their push-aware distance to the current target, producing a distance-aware, albeit suboptimal, assignment.
The resulting assignment is then refined by a local search that repeatedly performs pairwise swaps of target assignments whenever doing so reduces the maximum push-aware target--box distance.
The final assignment is denoted by \mtb.
Note that this refinement is guaranteed to terminate after finitely many iterations, since each accepted swap strictly decreases the objective.

\subsection{Agent--Stand Assignment}
Next, using the push-aware distance $\dist_p$, we identify a set of stands $S \subseteq V \setminus \bigcup_{b_j\in B}Q(b_j)$, where each stand represents a location from which an agent can push a box towards its assigned target, following \mtb.
A candidate stand $v \in V$ is discarded if it is occupied by another box or if the corresponding next box position is occupied by another box, since in either case the box cannot be pushed immediately.
Furthermore, certain push actions immediately lead to dead-end configurations.
For example, pushing a box may result in two adjacent boxes being trapped against a wall, rendering both boxes unpushable thereafter.
Such stand candidates are filtered out during the stand identification phase.

We then seek an agent--stand assignment \mas that minimises the maximum box-aware distance $\dist_{G\setminus B}$,
\[\max_{a_i\in A}\dist_{G\setminus B}(Q(a_i),\mas(a_i)),\]
analogous to the target--box assignment.
Unlike the previous step, however, a single box may admit multiple stands, whereas assigning multiple agents to the same box is redundant.
We thus adopt a constrained greedy assignment that processes agents sequentially.
Each agent is assigned to its unassigned nearest stand in terms of the box-aware distance, while ensuring that each box is assigned to at most one agent.
This is followed by pairwise-swap refinement, analogous to the construction of \mtb, to further reduce the maximum assigned distance, resulting in the final assignment \mas.

\subsection{Search Improvements}
\label{sec:algo:improvement}
The preceding sections describe the essential components of Sokoban-LaCAM.
Here, we present several search techniques that further improve its performance.
The techniques below include those specific to the anytime variant (i.e., \emph{\ao{Sokoban-LaCAM}}), which continues searching after the first feasible solution is found.

\paragraph{Random Restart} is a common technique for escaping search stagnation by non-deterministically returning to the initial search node~\cite{kautz2002dynamic}.
It has been successfully applied to several MAPF planners, including LaCAM itself~\cite{okumura2023improving}.
We adopt the same strategy by re-inserting the initial search node into \open with a small probability (e.g., $0.001$) whenever the search encounters a previously explored configuration at \cref{algo:lacam:check}.
Furthermore, after a feasible solution is found, random restarts are also triggered stochastically to encourage exploration of different search branches.

\paragraph{Admissible Heuristic} maps a configuration $Q$ to a lower bound on the cost required to reach a goal configuration.
Within \ao{LaCAM}, such heuristics are used for branch-and-bound, pruning search nodes that cannot improve upon the current best solution.
The trivial heuristic $h\sub{zero}(\cdot) \defeq 0$ is admissible, whereas tighter lower bounds substantially improve the convergence towards the optimum.
We therefore introduce the following heuristic, which combines lower bounds on the required target--box and agent--box movements:
\begin{align*}
  h\sub{T-B-A}(Q) =
  \max_{b_j \in B'}
  \biggl\{
  \min_{t \in T} \dist_p(t, b_j) +
  \max\left(\min_{a_i \in A} \dist_G(Q(a_i), b_j) - 1, 0\right)
  \biggr\},
\end{align*}
where $B'{=}\{b_j{\in}B \mid Q(b_j) \not\in T \}$ denotes a set of targets not on any target, and $\dist_G$ is the shortest path distance on $G$.

\paragraph{Search Stuck Detection} is implemented using the heuristic $h\sub{T-B-A}$.
Specifically, we monitor the search progress through this cost-to-go estimate.
If the heuristic value does not decrease for $d\sub{stuck}$ consecutive search iterations ($64$ in our experiments), we regard the search as being stuck, following the intuition of heuristic plateau detection~\cite{cohen2018local}, since successful search typically reduces this estimate over time.
In such cases, Sokoban-LaCAM performs a random restart, which improves solution discovery on challenging instances.
To preserve completeness in theory, $d\sub{stuck}$ should gradually increase over time.
However, our preliminary experiments showed no practical benefit from such a strategy, and we therefore use a fixed value of $d\sub{stuck}$ in default throughout the search.

\paragraph{Dead-end Configuration Pruning} can be applied immediately after configuration generation at \cref{algo:lacam:config-generator}.
For example, a configuration can be discarded if four boxes form a $2\times2$ block, rendering all of them permanently unpushable, unless all four cells are targets.
Another simple criterion compares the numbers of boxes and targets adjacent to the outer walls of the grid, allowing certain infeasible configurations to be detected efficiently.
Many additional dead-end patterns have been studied for Sokoban, and incorporating them is expected to further improve the planner.
Since they are primarily engineering optimisations, we leave a comprehensive treatment to future work.

\paragraph{Duplicate Detection} can be strengthened by exploiting the \emph{unlabelled} nature of Multi-Agent Sokoban.
In labelled settings, where individual agents and boxes are distinguished, two configurations $Q$ and $Q'$ are considered identical if $Q(o)=Q'(o)$ for every object $o\in A\cup B$.
In contrast, Multi-Agent Sokoban treats both agents and boxes as indistinguishable.
Accordingly, two configurations are regarded as identical when $Q(A)=Q'(A)\land Q(B)=Q'(B)$, where $Q(O)\defeq\bigcup_{o\in O}Q(o)$ denotes the set of locations occupied by objects in $O$.
This substantially reduces the number of distinct configurations stored in \explored, thereby accelerating anytime refinement.

\section{Evaluation}
We evaluate Sokoban-LaCAM from three perspectives:
\begin{itemize}
\item the benefit of introducing an MAPF-based search backbone through small-scale instances (\cref{sec:eval:ida});
\item the planner's ability to solve diverse instances with different map sizes, obstacle layouts, and numbers of agents and boxes (\cref{sec:eval:main}); and
\item the contribution of each technical component (\cref{sec:eval:ablation}).
\end{itemize}
All experiments are conducted on a desktop PC equipped with an Apple M1 Ultra CPU (20 cores) and \SI{64}{\giga\byte} RAM, with a maximum of 16 different runs in parallel using multi-threading.
Each instance is generated by randomly placing agents and boxes on a given grid map while excluding trivially infeasible cases; nevertheless, unsolvable instances may still be included.
Example instances are illustrated in \cref{fig:main}.
Our implementation is written in C++ and builds upon the LaCAM codebase for MAPF~\cite{okumura2023improving}.

{
\begin{table}[t!]
  \centering
  \small
  \setlength{\tabcolsep}{1.0mm}
  \begin{tabular}{rrrrrrr}
    \toprule
    \multicolumn{2}{c}{success (\% $\uparrow$)}
    & \multicolumn{2}{c}{time (\si{\milli\second} $\downarrow$)}
    & \multicolumn{3}{c}{sub-opt of \ao{LaCAM}}
    \\
    \cmidrule(lr){1-2}
    \cmidrule(lr){3-4}
    \cmidrule(lr){5-7}
    \ao{IDA}
    & \ao{LaCAM}
    & \ao{IDA}
    & \ao{LaCAM}
    & init
    & \SI{30}{\second}
    & opt-proven
    \\\midrule
    30.0
    & 93.3
    & 3117{\tiny$\pm$2663}
    & <\SI{0.1}{\milli\second}
    & 1.44
    & 1.00
    & 24/27${=}$89\%
    \\\bottomrule
  \end{tabular}
  \caption{
    Performance of \ao{Sokoban-LaCAM} against \ao{IDA} on the 90 instances of \mapname{empty-5-5} from \cref{fig:main} under a \SI{30}{\second} time limit.
    Among them, both planners solve 27 instances.
    For these commonly solved instances, we report the time required by \ao{Sokoban-LaCAM} to obtain its initial solutions, together with the suboptimality factors for initial and final solutions using the optimal \ao{IDA} solutions as references.
    Notably, \ao{Sokoban-LaCAM} finds initial solutions for all 27 instances almost instantly, and all of them eventually converge to the optimum.
    Furthermore, \ao{Sokoban-LaCAM} proves optimality for most of these instances (see \emph{opt-proven}).
  }
  \label{table:ida}
\end{table}
}

{
\newcommand{\fheight}{0.16\linewidth}
\newcommand{\fwidth}{0.16\linewidth}
\newcommand{\entry}[3]{
  \begin{scope}[xshift=#1*0.16\linewidth,yshift=-#2*0.352\linewidth]
    \node[anchor=south] at (\fwidth * 0.6, -0.01\linewidth) {\scriptsize \mapname{#3}};
    \node[anchor=north west] at (0, 0)
         {\includegraphics[width=\fwidth,height=\fheight]{fig/raw/heatmaps/#3_pibt_rollout_no_stop_at_duplicate_paper}};
    \node[anchor=north west] at (0, -\fheight)
         {\includegraphics[width=\fwidth,height=\fheight]{fig/raw/heatmaps/#3_lacam_paper}};
  \end{scope}
}

\begin{figure*}[h!]
\centering
\begin{tikzpicture}
  \scriptsize
  \entry{0}{0}{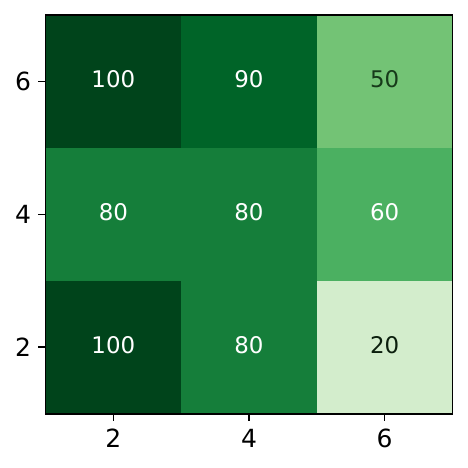}
  \entry{1}{0}{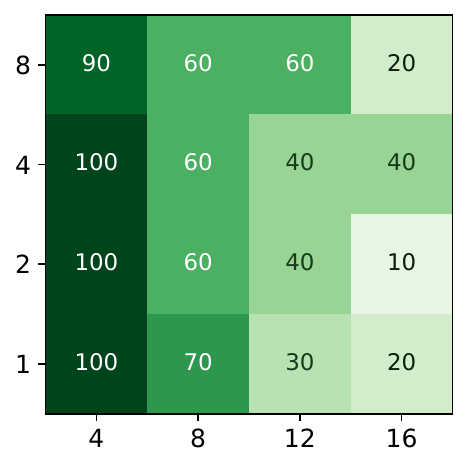}
  \entry{2}{0}{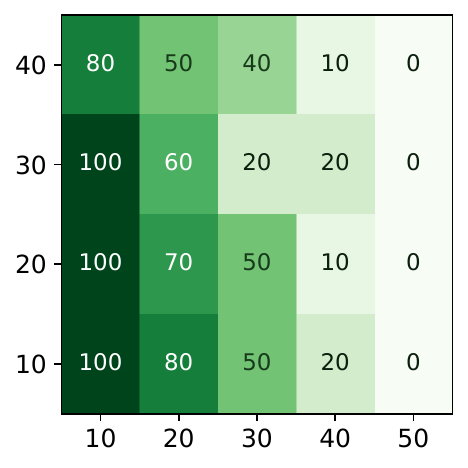}
  \entry{3}{0}{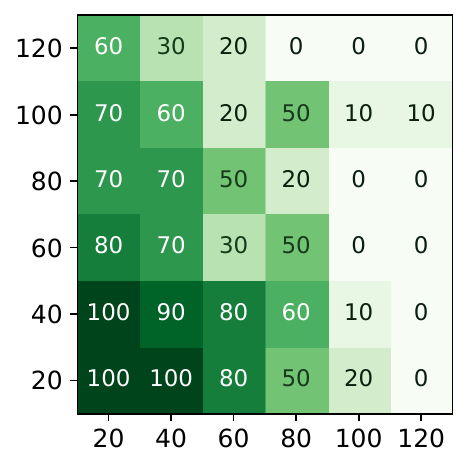}
  \entry{4}{0}{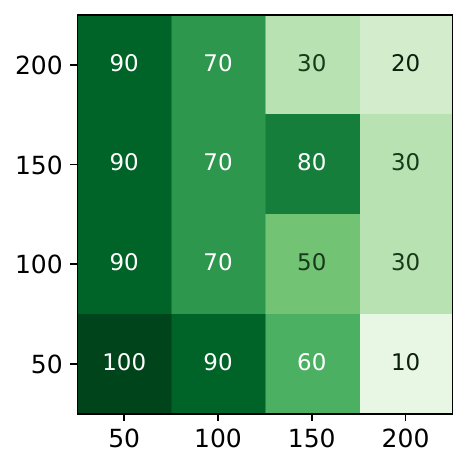}
  \entry{0}{1}{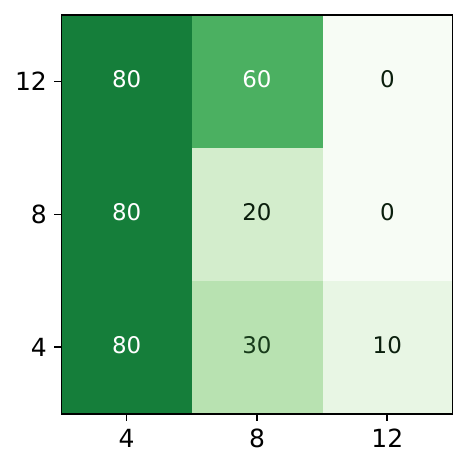}
  \entry{1}{1}{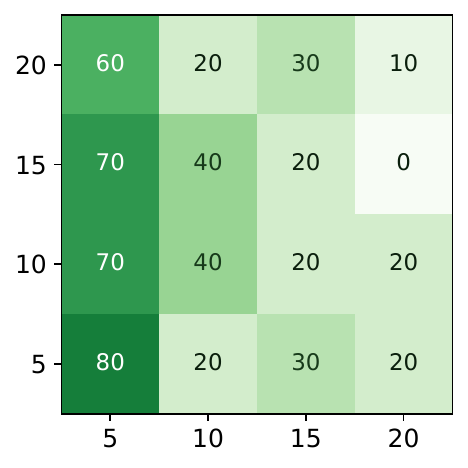}
  \entry{2}{1}{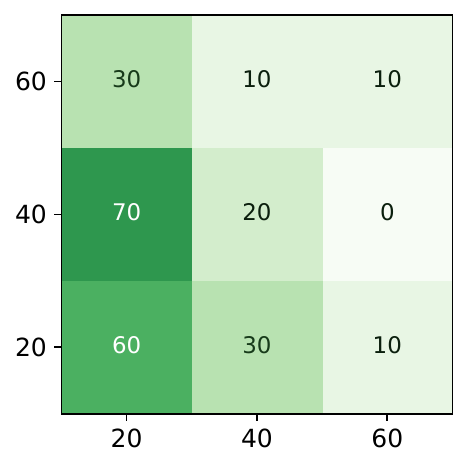}
  \entry{3}{1}{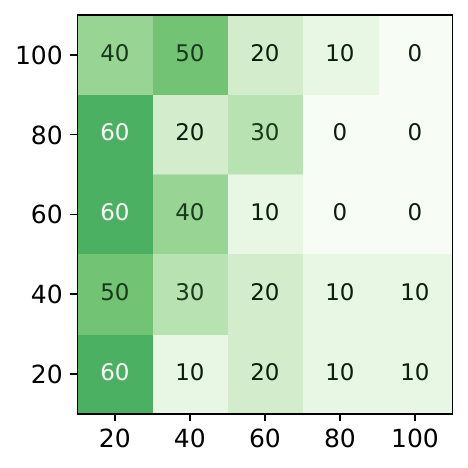}
  \entry{4}{1}{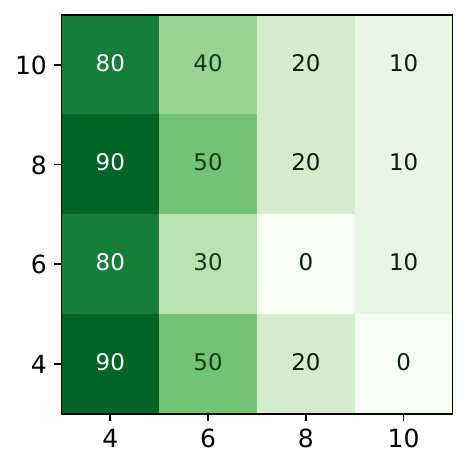}
  \entry{5}{1}{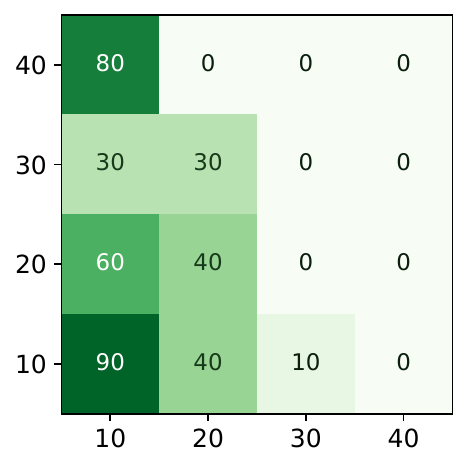}
  %
  \begin{scope}[xshift=5.2*0.16\linewidth,yshift=0]
    \node[anchor=north west] at (0, 0)
         {\includegraphics[height=0.31\linewidth]{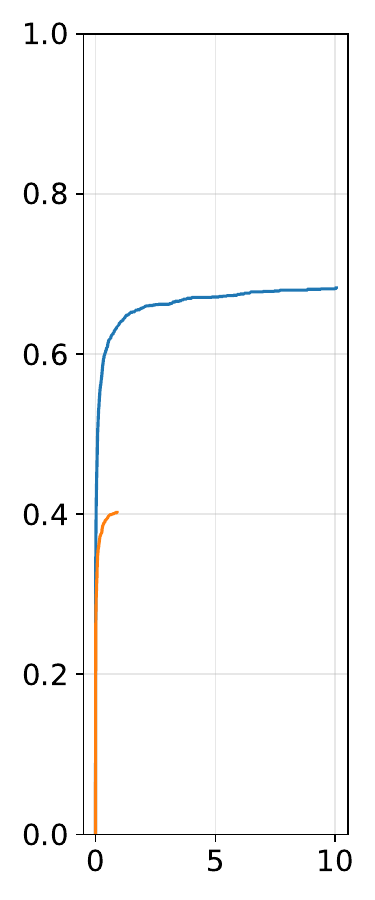}};
    \node[rotate=90] at (0, -\fheight*1) {success rate (of 1{,}880 instances)};
    \node[] at (\fwidth*0.53, -\fheight*2) {runtime [\si{\second}]};
    \node[anchor=west] at (\fwidth*0.25, -\fheight*1.2) {\textcolor[HTML]{ff7f0e}{Sokoban-}};
    \node[anchor=west] at (\fwidth*0.25, -\fheight*1.3) {\textcolor[HTML]{ff7f0e}{PIBT {\tiny (40.2\%)}}};
    \node[anchor=west] at (\fwidth*0.20, -\fheight*0.5) {\textcolor[HTML]{1f77b4}{Sokoban-}};
    \node[anchor=west] at (\fwidth*0.20, -\fheight*0.6) {\textcolor[HTML]{1f77b4}{LaCAM {\tiny (68.2\%)}}};
    \node[] at (\fwidth*0.5, 0) {\textbf{Aggregated}};
  \end{scope}
  %
  \node[] at (\linewidth*0.5, 0.35) {\small\textbf{success [\%]}};
  \node[] at (\linewidth*0.5, -\fheight*2.07) {\textbf{\#Boxes}};
  \node[] at (\linewidth*0.5, -\fheight*4.28) {\textbf{\#Boxes}};
  \node[rotate=90] at (-0.02*\linewidth, -\fheight*1) {\textbf{\#Agents}};
  \node[rotate=90] at (-0.02*\linewidth, -\fheight*3.25) {\textbf{\#Agents}};
  \node[rotate=90] at (-0.0*\linewidth, -\fheight*0.5) {Sokoban-PIBT};
  \node[rotate=90] at (-0.0*\linewidth, -\fheight*1.5) {Sokoban-LaCAM};
  \node[rotate=90] at (-0.0*\linewidth, -\fheight*2.75) {Sokoban-PIBT};
  \node[rotate=90] at (-0.0*\linewidth, -\fheight*3.75) {Sokoban-LaCAM};
  %
  \begin{scope}[xshift=0,yshift=\fheight*1.15]
    \tiny
    \newcommand{\img}[4]{
      \node[anchor=north] at (#1*\fwidth, 0)
         {\includegraphics[height=0.12\linewidth,clip,trim={0.5cm 0.5cm 0.5cm 1.4cm}]{fig/raw/instances/#2_#3a#4b}};
      \node[] at (#1*\fwidth, 0.02\linewidth) {\mapname{#2}};
      \node[] at (#1*\fwidth, 0.005\linewidth) {$|A|{=}#3, |B|{=}#4$};
    }
    \img{0.4}{empty-8-8}{4}{16}
    \img{1.32}{random-8-8-10}{10}{10}
    \img{2.1}{random-16-16-10}{20}{20}
    \img{2.87}{random-32-32-10}{60}{60}
    \img{3.64}{random-64-64-10}{100}{100}
    \img{4.65}{warehouse_c3-1}{10}{10}
    \img{5.6}{warehouse_c4-2}{30}{30}
    \node[anchor=west] at (0, -0.86*\fheight)
         {
           $\bullet$ agent\quad
           \textcolor[RGB]{172,138,104}{$\blacksquare$} box\quad
           \textcolor[RGB]{126,103,77}{$\Box$} target
         };
  \end{scope}
\end{tikzpicture}
\caption{
  Planning performance of Sokoban-LaCAM and its underlying configuration generator, Sokoban-PIBT.
  The top row illustrates representative problem instances.
  The bottom rows report the success rate under a \SI{10}{\second} time limit across different maps and varying numbers of agents and boxes.
  Each scenario consists of ten instances.
  The centre-rightmost plot shows the aggregated success rate as a function of elapsed planning time.
}
\label{fig:main}
\end{figure*}
}

{
\newcommand{\head}[1]{
  \multicolumn{2}{c}{\begin{tabular}{c}#1\end{tabular}}
}
\newcommand{\ci}[1]{{\tiny{$\pm$}#1}}
\newcommand{\w}[1]{\textbf{#1}}
\begin{figure*}[h!]
  \centering
  \begin{tikzpicture}

  \node[anchor=north west] at (0, 0) {
  \footnotesize
  \setlength{\tabcolsep}{1.0mm}
  \begin{tabular}{lrrrlrlrlrrrrr}
    \toprule
    & \head{success\\(rate $\uparrow$)}
    & \head{init time\\(\si{\milli\second} $\downarrow$)}
    & \head{init cost\\impr (\% $\uparrow$)}
    & \head{\SI{10}{\second} cost\\impr (\% $\uparrow$)}
    & \head{\#search\\nodes}
    & \head{\#const-\\raints}
    \\\midrule
    Sokoban-LaCAM
    && \w{0.539}
    & 148 & \ci{87}
    & \w{0.0} & \ci{0.0}
    & \w{21.4} & \ci{2.0}
    && 291K
    && 2.6M
    \\
    $-$unlabelled duplicate
    && 0.537
    & 175 & \ci{106}
    & -2.0 & \ci{2.1}
    & 20.9 & \ci{2.1}
    && 332K
    && 2.8M
    \\
    $-$dead-end pruning
    && 0.532
    & 743 & \ci{88}
    & \w{0.1} & \ci{2.5}
    & \w{21.3} & \ci{2.0}
    && 347K
    && 3.0M
    \\
    $-$random restart
    && \w{0.539}
    & \w{131} & \ci{79}
    & -3.2 & \ci{3.4}
    & -0.9 & \ci{3.4}
    && 53K
    && 0.9M
    \\
    $-$stuck detection
    && 0.485
    & 180 & \ci{92}
    & -44.1 & \ci{30.2}
    & 20.9 & \ci{2.0}
    && 289K
    && 2.6M
    \\\bottomrule
  \end{tabular}
  };
  \begin{scope}[xshift=0.78\linewidth,yshift=-0.01\linewidth]
    \scriptsize
    \node[anchor=north west] at (0, 0)
       {\includegraphics[width=0.16\linewidth,height=0.16\linewidth]
         {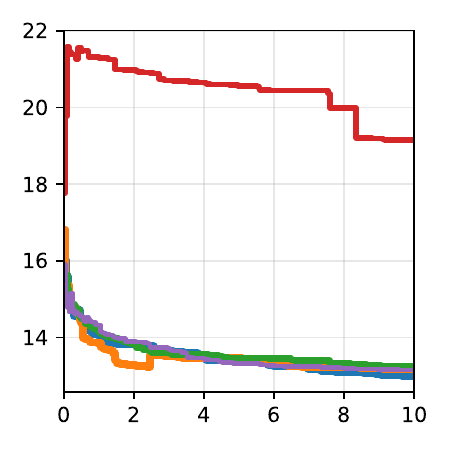}};
    \node[rotate=90] at (0, -0.08\linewidth) {makespan};
    \node[] at (0.09\linewidth, -0.165\linewidth) {time [\si{\second}]};
    \node[] at (0.09\linewidth, -0.005\linewidth) {\mapname{random-8-8-10}};
    %
    \definecolor{c1}{RGB}{194,59,49}
    \node[anchor=west] at (0.035\linewidth, -0.05\linewidth) {\tiny\textcolor{c1}{$-$random restart}};
    \definecolor{c1}{RGB}{83,156,56}
    \node[anchor=west] at (0.035\linewidth, -0.08\linewidth) {\tiny\textcolor{c1}{$-$dead-end pruning}};
    \definecolor{c1}{RGB}{142,108,186}
    \node[anchor=west] at (0.035\linewidth, -0.09\linewidth) {\tiny\textcolor{c1}{$-$stuck detection}};
    \definecolor{c1}{RGB}{236,134,48}
    \node[anchor=west] at (0.035\linewidth, -0.10\linewidth) {\tiny\textcolor{c1}{$-$unlabelled duplicate}};
    \definecolor{c1}{RGB}{64,119,177}
    \node[anchor=west] at (0.035\linewidth, -0.11\linewidth) {\tiny\textcolor{c1}{Sokoban-LaCAM}};
  \end{scope}

  \end{tikzpicture}
  \caption{
    Ablation study and anytime performance of \ao{Sokoban-LaCAM} on the \mapname{random-*} maps from \cref{fig:main}.
    Except for the \emph{success} rate under the \SI{10}{\second} time limit, the left table reports averages over the 274 instances solved by all variants for a fair comparison, together with 95\% confidence intervals.
    \emph{init time} denotes the time required to find the first feasible solution.
    \emph{init cost impr} reports the makespan improvement relative to Sokoban-LaCAM upon finding the initial solution.
    Similarly, \emph{\SI{10}{\second} cost impr} reports the relative makespan improvement at the time limit.
    We additionally report the number of search nodes generated by LaCAM (\emph{\#search nodes}) and the number of generated constraints (\emph{\#constraints}).
    The right plot focuses on the average makespan over planning time on \mapname{random-8-8-10}, computed from the 72 instances solved by all variants.
  }
  \label{fig:ablation}
\end{figure*}
}

\subsection{Comparison with \ao{IDA}}
\label{sec:eval:ida}

Single-agent Sokoban has long been tackled via \ao{IDA}~\cite{korf1985depth}, together with domain-specific enhancements~\cite{junghanns2001sokoban}.
We therefore adapt this approach to Multi-Agent Sokoban as a natural baseline.
Our \ao{IDA} implementation performs makespan-optimal search over configurations while incorporating the admissible heuristic $h\sub{T-B-A}$ and the dead-end pruning techniques introduced in \cref{sec:algo:improvement}.
We then compare \ao{IDA} with \ao{Sokoban-LaCAM}, an anytime search that continues the search towards makespan-optimal solutions.
Both planners are evaluated on the obstacle-free $5\times5$ map, denoted by \mapname{empty-5-5}, with the number of agents $n \in \{2,4,6\}$ and boxes $m \in \{2,4,6\}$.
Each setting contains 10 instances, resulting in 90 instances in total.
The planning time limit is set to \SI{30}{\second}.

\Cref{table:ida} presents the aggregated results.
Overall, at this problem scale, \ao{Sokoban-LaCAM} solves more instances, finds initial feasible solutions almost instantly, and eventually converges to optimal solutions on all instances solved by \ao{IDA}, proving optimality for most of them.
The substantially lower success rate of \ao{IDA} suggests that, while admissible heuristics are essential for systematic search, $h\sub{T-B-A}$ alone is insufficient to guide the search effectively in Multi-Agent Sokoban.
This highlights the importance of configuration generator--based search combined with depth-first exploration.
Rather than relying primarily on increasingly informative heuristics to evaluate configurations, LaCAM directly generates promising successor configurations without \emph{seeing} unpromising ones.
This design leads to substantially more efficient search in the vast configuration space.

\subsection{Stress Test}
\label{sec:eval:main}
We next stress-test Sokoban-LaCAM across a wide range of problem settings, as summarised in \cref{fig:main}.
The benchmark includes obstacle-free maps, maps with 10\% randomly distributed obstacles, and warehouse-like environments;
some of them are from the MAPF benchmark~\cite{stern2019def}.
Since \ao{IDA} largely fails at this scale, it is omitted from the following evaluation.
Instead, we additionally evaluate the underlying configuration generator, Sokoban-PIBT, by repeatedly applying it from the current configuration until all boxes reach their targets or the makespan upper bound is exceeded.
The planning time limit is set to \SI{10}{\second}.
Each scenario consists of 10 instances.
Note that our goal is not to solve highly challenging puzzle instances, but rather to assess the planner's ability to coordinate multiple agents.

The results in \cref{fig:main} reveal that Sokoban-PIBT alone can solve relatively easy instances, while the systematic search framework of Sokoban-LaCAM substantially improves the overall planning performance.
Nevertheless, Sokoban-LaCAM struggles more on instances with many boxes than many agents, suggesting that successful box manipulation requires more careful long-horizon reasoning to avoid irreversible dead-ends.
Despite this, Sokoban-LaCAM successfully solves instances with more than 100 agents or more than 100 boxes, far beyond the reach of conventional search schemes, demonstrating that recent advances in MAPF provide a powerful foundation for scalable multi-agent planning beyond canonical MAPF.
We additionally report that replacing the proposed preference construction strategy, which performs target--box and agent--stand assignment, with random preference construction solves only 2.8\% of instances, highlighting the importance of informed preference construction.

\subsection{Ablation Study}
\label{sec:eval:ablation}

Finally, \cref{fig:ablation} investigates the contribution of each search component introduced in \cref{sec:algo:improvement} beyond the basic LaCAM framework.
The evaluation uses the instances on the \mapname{random-x-x-10} maps, where $x \in \{8,16,32,64\}$.

Among the proposed techniques, search stuck detection plays the most important role in solving difficult instances.
Meanwhile, random restart has a substantial impact on solution quality, as also evidenced by the anytime behaviour shown in the right plot.
Without it, the depth-first nature of LaCAM tends to continue digging into a single search branch, resulting in longer makespans and poorer exploration of the search space, as also reflected by the smaller numbers of generated search nodes and search constraints.
The remaining components also contribute consistently to the overall performance.
Although unlabelled duplicate detection and dead-end pruning introduce additional computation per node, their removal results in more generated search nodes and constraints, indicating that the extra overhead is more than compensated for by more effective pruning.

\section{Discussion}
\label{sec:discussion}
We presented Sokoban-LaCAM for Multi-Agent Sokoban, a largely unexplored planning problem that tightly integrates task assignment, object manipulation, and coordinated path planning.
Our experiments show that Sokoban-LaCAM can efficiently solve large-scale instances.

More importantly, however, our objective is not to solve Sokoban itself.
Rather, we use it as a testbed for investigating whether recent advances in MAPF can be transferred beyond the canonical MAPF formulation.
The results suggest that configuration-based MAPF search, exemplified by the combination of PIBT and LaCAM, provides a promising foundation for more general multi-agent planning problems.
This perspective is particularly encouraging because conventional search techniques often struggle to scale to multi-agent settings due to the rapidly growing joint search space, whereas specialised MAPF search backbones can be extended with relatively modest domain-specific modifications, as demonstrated in this work.
In this regard, we believe that MAPF should increasingly be viewed not merely as a standalone planning problem, but as a scalable algorithmic primitive upon which more sophisticated multi-agent planning systems can be built.
With this in mind, a promising future direction is to integrate learned heuristics into Sokoban-LaCAM, following recent successes in learning-guided LaCAM for challenging problems~\cite{jain2026graph}.

\section*{Acknowledgments}
This research was conducted by AIST and under FRONTia, a Japanese national program led by the Ministry of Economy, Trade and Industry of Japan (METI) and the New Energy and Industrial Technology Development Organization (NEDO), aimed at developing a domestic multimodal foundation model for AI robots and physical AI.

\bibliographystyle{unsrtnat}
\bibliography{sty/ref-macro,ref}
\end{document}